\documentclass[11pt]{article}
\usepackage[utf8]{inputenc}
\usepackage[T1]{fontenc}
\usepackage{textcomp}
\usepackage{gensymb}
\usepackage{amsmath}
\usepackage{relsize}
\usepackage{lmodern}
\usepackage{slantsc}
\usepackage{geometry}                % See geometry.pdf to learn the layout options. There are lots.
\usepackage{graphicx}
\usepackage{amssymb}
\usepackage{amsmath}
\usepackage{mathtools}
\usepackage{epstopdf}
\usepackage{subfig}

\usepackage[font={footnotesize,bf}]{caption}

\usepackage[affil-it]{authblk}

\title{Ultra-Efficient Thermophotovoltaics Exploiting Spectral Filtering by the Photovoltaic Band-Edge}
\author{Vidya Ganapati, T. Patrick Xiao, Eli Yablonovitch}
\affil{University of California, Berkeley}
\begin{document}
\maketitle

\section{Abstract}
Thermophotovotaics convert thermal radiation from local heat sources to electricity. A new breakthrough in creating highly efficient thin-film solar cells can potentially enable thermophotovoltaic systems with unprecedented high efficiency. The current 28.8\% single-junction solar efficiency record, by Alta Devices, was achieved by recognizing that a good solar cell needs to reflect infrared band-edge radiation at the back surface, to effectively recycle infrared luminescent photons. The effort to reflect band-edge luminescence in solar cells has serendipitously created the technology to reflect all infrared wavelengths, which can revolutionize thermophotovoltaics. We have never before had such high back reflectivity for sub-bandgap radiation, permitting step-function spectral control for the first time. Thus, contemporary efficiency advances in solar photovoltaic cells create the possibility of realizing a $>50\%$ efficient thermophotovoltaic system.

\section{Introduction}
\begin{figure}[htbp]
\includegraphics[scale=0.5]{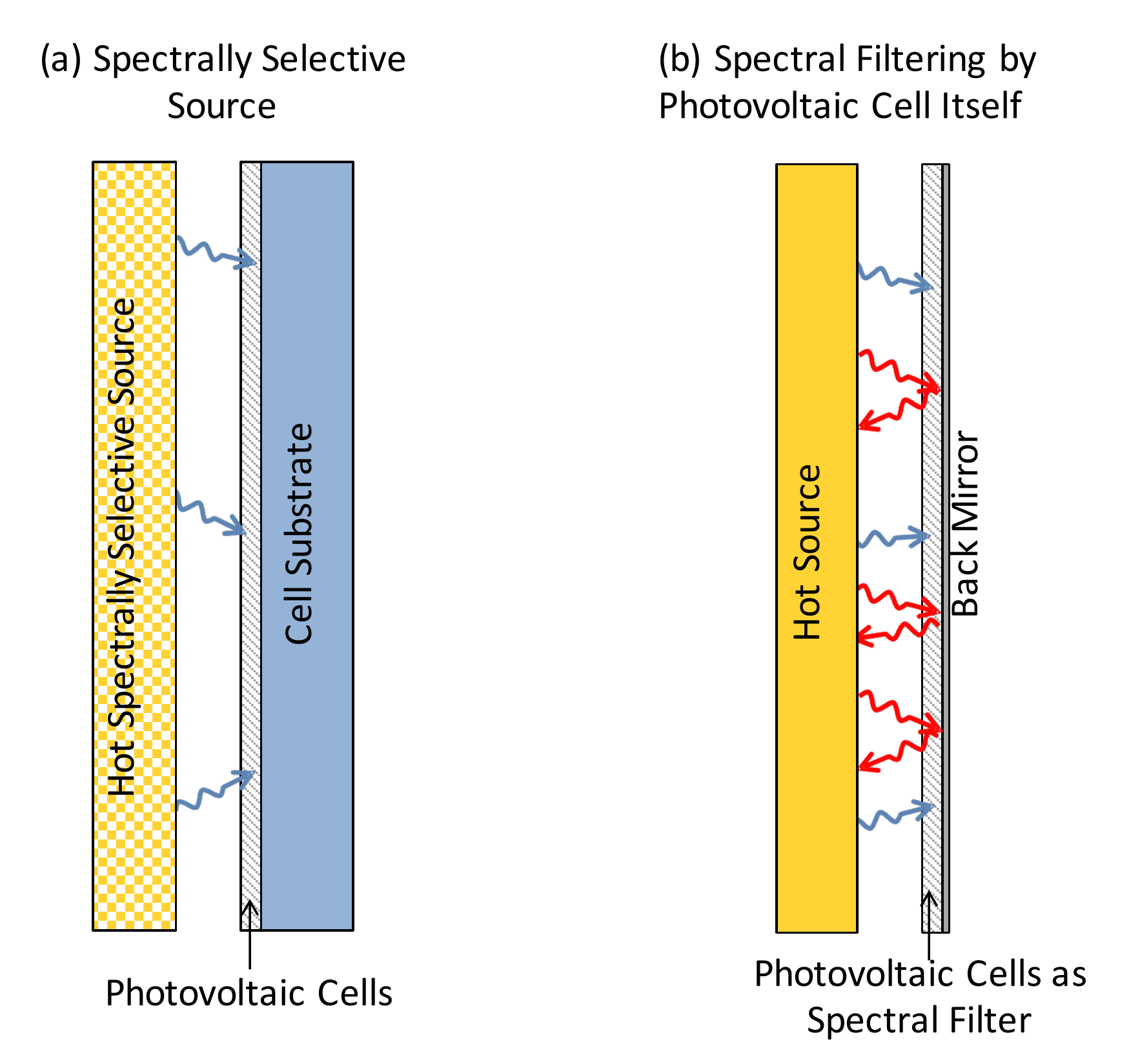}
\centering
\caption{For high efficiency, below bandgap photons need to be recycled back to the hot source. This is done with (a) a spectrally selective source, such as a photonic crystal, which exhibits low emissivity of below bandgap photons (low emissivity of photons is analogous to high reflectivity of photons back to the source) or (b) with a mirror on the back of the photovoltaic cells which reflects the infrared photons back to the source.}
\label{fig1}
\end{figure}

In a photovoltaic cell, thermal radiation is converted to electricity. In solar photovoltaics, the thermal radiation comes from the sun, an approximate blackbody at $5500\degree$C, $1.5 \times 10^8$ km away from Earth. In thermophotovoltaics, the thermal radiation can come instead from a local hot source. The hot source can be generated from combustion of fuel \cite{fraas_tpv_2007}, concentrated sunlight \cite{harder_theoretical_2003}, or a nuclear power source \cite{teofilo_thermophotovoltaic_2008}. Photons radiate from the hot source, with the radiation spectrum depending on the temperature and material properties. As these sources are generally much cooler than the sun, the emitted thermal radiation will be mainly composed of very low energy photons, unusable by a photovoltaic cell. In order to efficiently convert from heat to electricity, low bandgap photovoltaic cells are needed, as well as a spectral filter to recycle the large number of very low energy photons back to the source. Ideally, the spectral filter needs to allow above-bandgap photons to be absorbed by the photovoltaic cell, while reflecting the below-bandgap photons back to the source. The source can be in the spectral filter itself; low emissivity of certain photon energies is analogous to having a high reflectivity of those photons back to the source. In this case, the emissivity spectrum needs to match the absorptivity spectrum of the photovoltaic cell. This approach to spectral filtering is shown in Fig.~\ref{fig1}a, where a photonic crystal, which is optimized for low emissivity below the cell bandgap and high emissivity above the cell bandgap, acts as the source.

It was recognized many years ago that the semiconductor band-edge itself could provide excellent spectral filtering for thermophotovoltaics, providing that the unused below bandgap radiation can be efficiently reflected back to the heat source \cite{werth_thermo-photovoltaic_1967}. The semiconductor band-edge can allow the above-bandgap photons to be absorbed in the cell, and the below-bandgap photons to be transmitted through the cell. If the cell has a highly reflective back mirror, the below-bandgap photons get reflected back to the hot source. For a spectrally selective source, it is a large challenge to match the emissivity of the source to the semiconductor band-edge of the photovoltaic cell. On the other hand, when using the photovoltaic band-edge itself as the spectral filter, band-edge alignment is achieved by default. Fig.~\ref{fig1} schematically shows the difference between using a selective emitter, such as a photonic crystal, for the spectral filter, and using the photovoltaic band-edge itself as the spectral filter. Fig.~\ref{fig1}a shows a traditional photovoltaic cell that is grown on a substrate with no back mirror, as the spectral control is solely in the source. In Fig.~\ref{fig1}b, we show a photovoltaic cell with the substrate removed and replaced by a highly reflective back mirror.

Using the photovoltaic band-edge as the spectral filter puts a burden on the infrared reflectivity of photovoltaic cells toward their unused radiation. In the design of a conventional solar cell, this infrared radiation is typically ignored. A new breakthrough in record-breaking efficient thin-film solar cells has changed the situation. The current 28.8\% single-junction solar efficiency record, by Alta Devices, was achieved by recognizing that a good solar cell needs to have high back mirror reflectivity to allow internal luminescence to escape from the front surface of the cell \cite{miller_strong_2012}. At the maximum power point of a solar cell, a small percentage of photons are absorbed and not collected as current, but instead re-emitted internally in the cell. In order to obtain a high cell voltage, these re-emitted internal photons, which can be re-emitted with energies below the band-edge, must make it out of the front surface of the solar cell \cite{miller_strong_2012}. Thus it is important to have high back reflectivity of internal infrared band-edge radiation, to effectively recycle them out the front surface \cite{miller_strong_2012}. For high back reflectivity, it is essential to remove the original semiconductor substrate, which absorbs infrared luminescence, and to replace it with a highly reflective mirror. The solar cell efficiency record crept up as the back reflectivity behind the photovoltaic film was increased, from 96\% reflectivity, to 97\%, to finally 98\% luminescent reflectivity; each produced a new world efficiency record \cite{green_solar_2011-1,green_solar_2011,green_solar_2012,green_solar_2012-1}.

\begin{figure}[htbp]
\includegraphics[scale=0.4]{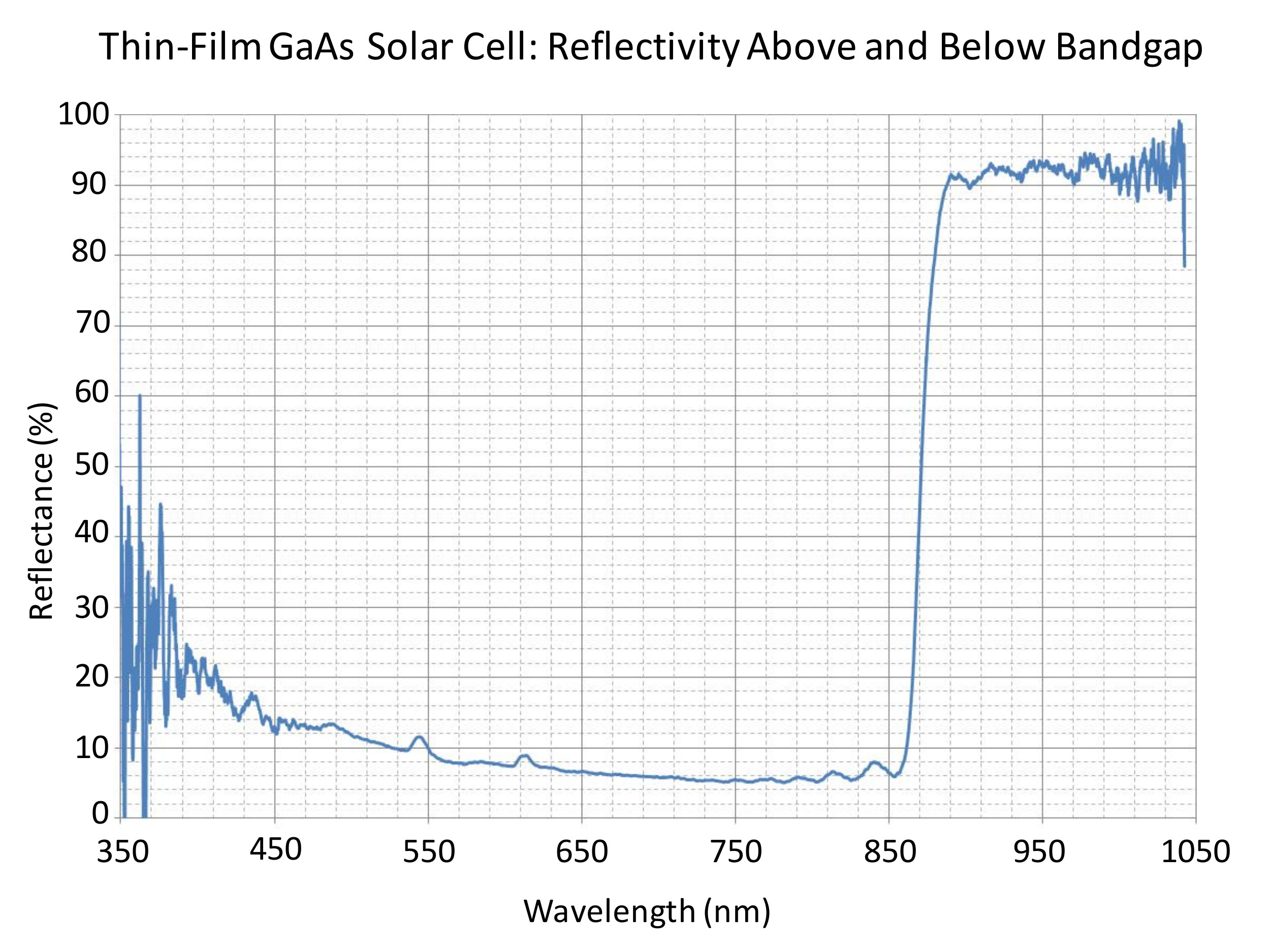}
\centering
\caption{Reflectivity of a standard production GaAs solar cell, courtesy of Alta Devices Inc. (the world record cell had an even higher sub-bandgap reflectivity of $\approx 98\%$).  Achieving the same step function response, in an InGaAs alloy, would be ideal for thermophotovoltaics.}
\label{fig2}
\end{figure}

The effort to reflect band-edge luminescence in solar cells has serendipitously created the technology to reflect all infrared wavelengths, which can revolutionize thermophotovoltaics. Fig.~\ref{fig2} shows the reflectance as a function of wavelength for a standard production Alta Devices Gallium Arsenide (GaAs) solar cell. The high back reflectivity is >92\% for the sub-bandgap radiation, and we see a clear step function at the bandgap of $\approx 870$ nm. 

The principle of using the photovoltaic band-edge as the spectral filter for thermophotovoltaics was demonstrated in 1978, with a silicon solar cell and a hot source at $2000\degree$C degrees, achieving an efficiency of 26\% \cite{swanson_silicon_1978}. An efficiency of 20.6\% was achieved in 2003 with the same concept at a lower temperature of $\approx 1050\degree$C, using InGaAs cells with a 0.6 eV bandgap \cite{siergiej_20_2003}. However, the large thickness of the InP substrate on top of the back mirror led to relatively large free carrier absorption, degrading the reflectivity of the cell. The efficiency was improved to 23.6\% in 2004 \cite{wernsman_greater_2004}, with the addition of a multilayer dielectric filter on the front surface of the cell. These results demonstrate that high thermophotovoltaic efficiency is accessible by using the photovoltaic band-edge as a spectral filter.

Here, we propose to use thin films of direct bandgap, epitaxially grown semiconductors for thermophotovoltaics, to allow for high absorption of above-bandgap photons but low free carrier absorption. For the record-breaking GaAs cells, Alta Devices epitaxially grew films of GaAs on substrate, with the film separated from the substrate by a thin sacrificial layer of Aluminum Arsenide (AlAs). The AlAs sacrificial layer was then chemically etched, and the film was lifted off the substrate and placed on a reflective back mirror \cite{yablonovitch_extreme_1987, kayes_27.6_2011}. This epitaxial lift-off procedure is crucial to achieving high back mirror reflectivity, as it allows the substrate to be replaced by a back mirror. However, though GaAs photovoltaic cells are a well-developed technology with high efficiency under sunlight, the bandgap of GaAs (1.4 eV) is too large to obtain high efficiency under cooler source temperatures of $1200\degree$C to $1500\degree$C desirable for thermophotovoltaic systems. Nonetheless, with the same epitaxial lift-off process, we can create direct-band InGaAs cells with smaller bandgaps. This work will calculate the optimal bandgaps needed for thermophotovoltaics.

\section{Theory}
\begin{figure}[htbp]
  \centering
  \subfloat{\includegraphics[scale=0.5]{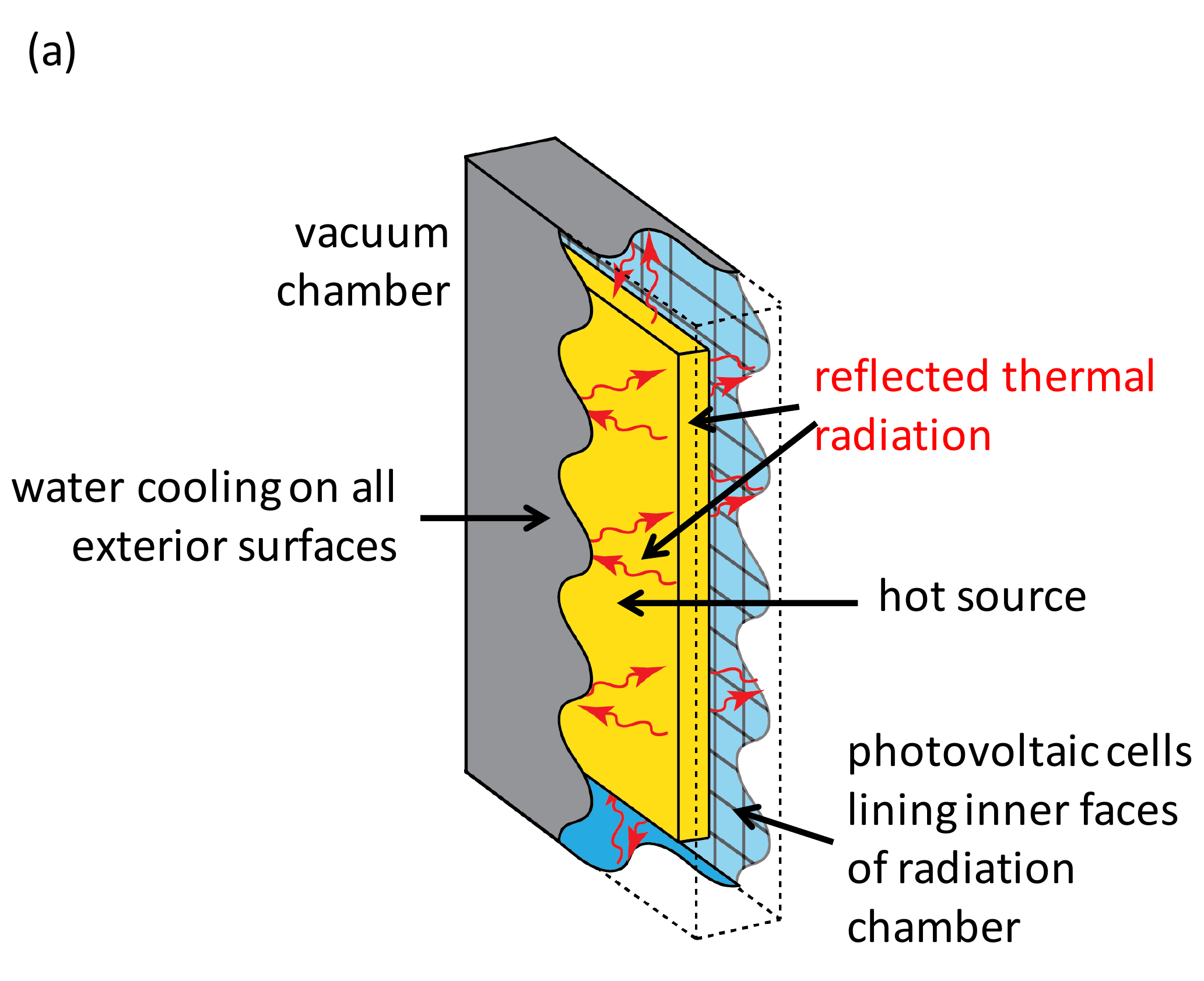}\label{fig3a}}
  \hfill
  \subfloat{\includegraphics[scale=0.5]{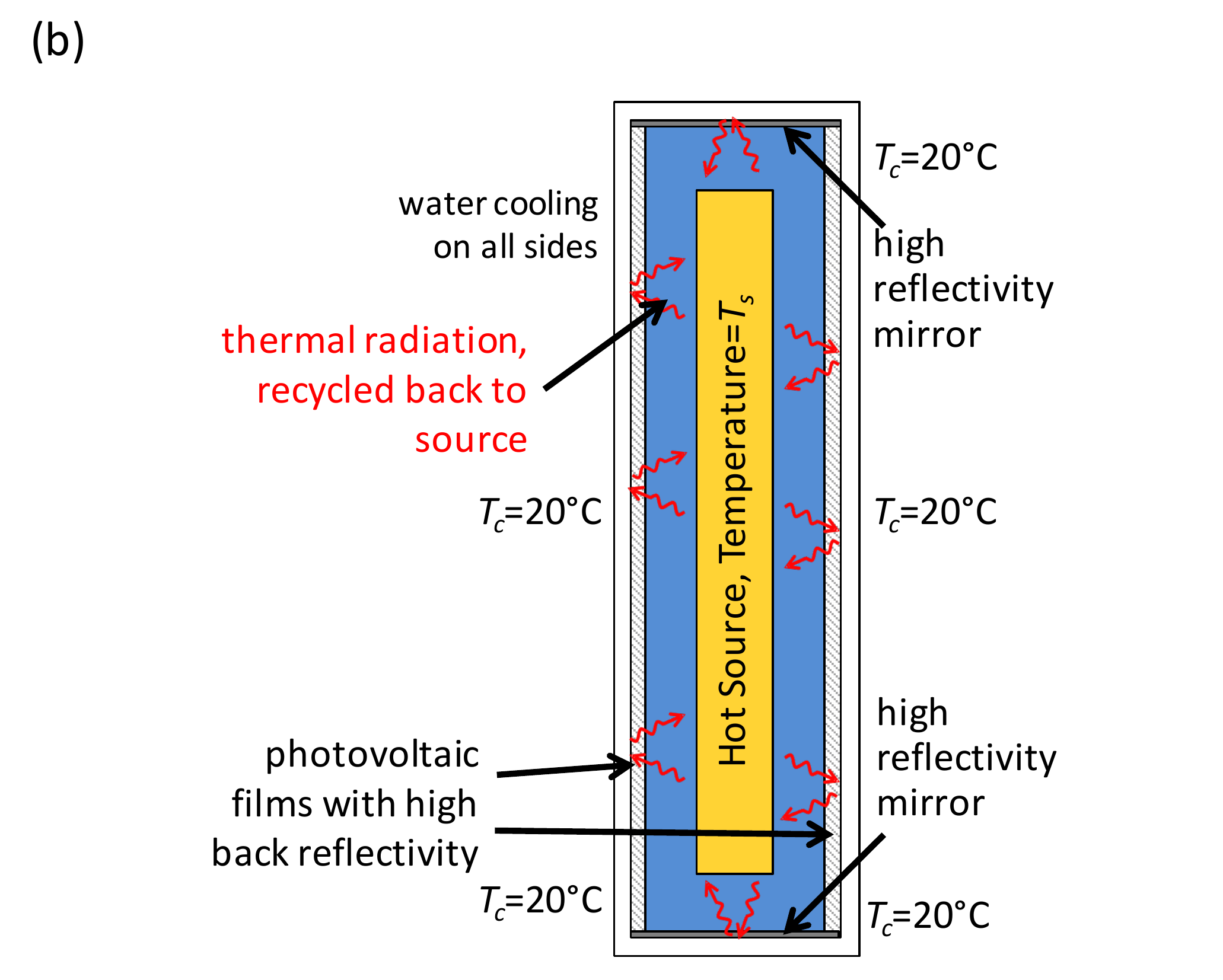}\label{fig3b}}
  \caption{(a) Perspective view and (b) cross-sectional view of the thermophotovoltaic chamber.}
  \label{fig3}
\end{figure}

We analyze the efficiency of thermophotovoltaic systems; an example system is diagrammed in Fig.~\ref{fig3}. A hot radiation source is enclosed by a cavity, the thermophotovoltaic chamber. Photovoltaic cells line the largest two inner faces of a thin rectangular cavity and are plane parallel to a hot radiation source. The other 4 inner faces of the cavity are lined with reflective mirrors. The interior of the cavity is under vacuum to minimize non-radiative heat transfer, and the hot radiation source emits thermal radiation as a blackbody at temperature $T_{source}=T_s$. All of the outside surfaces of the thermophotovoltaic chamber are water-cooled, maintaining the solar cells at a temperature $T_{cell}=T_c$. The photovoltaic cells have back mirrors with reflectivity as a function of photon energy. We denote the reflected sub-bandgap photons with red arrows.

We follow the formulation for solar cell efficiency, given in Refs. \cite{miller_strong_2012, shockley_detailed_1961} to derive the efficiency equation for thermophotovoltaics. The blackbody radiation from the hot source incident on the cold photovoltaic cells is given as:
\begin{equation}
b_s(E)=\frac{2 \pi E^2}{c^2 h^3 \left( \exp \left( \frac{E}{k_B T_s} \right)-1 \right)},
\label{eq1}
\end{equation}
where $b_s(E)$ is the blackbody radiation flux in units of photons/time/area/energy, $E$ is the photon energy, $c$ is the speed of light, $h$ is Planck's constant, $k_B$ is the Boltzmann constant, and $T_s$ is the temperature of the hot radiation source.

We assume the photovoltaic cells have step function absorption, absorbing all photon energies above the bandgap, $E_g$. The short circuit current density of the cells, $J_{sc}$, is given by:
\begin{equation}
J_{sc}= q \int_0^\infty a(E) b_s(E) dE = q \int_{E_g}^\infty b_s(E)dE,
\label{eq2}
\end{equation}
where $q$ is the charge of an electron and $a(E)$ is the absorption as a function of photon energy. As we assume step function absorption, the equation simplifies to the expression on the right.

In the dark, when not illuminated by the source, the photovoltaic cells emit blackbody radiation at temperature $T_c$. The radiation emitted from the cells in the dark, similar to Eqn.~\ref{eq1}, is given as:
\begin{equation}
b_c(E)=\frac{2 \pi E^2}{c^2 h^3 \left( \exp \left( \frac{E}{k_B T_c} \right)-1 \right)},
\label{eq3}
\end{equation}
where $b_c(E)$ is the blackbody radiation from the photovoltaic cells.

The cells have some external fluorescence yield of $\eta_{ext}$ (defined here, as in Ref. \cite{miller_strong_2012}, as the ratio of radiative recombination out the front surface of the cell to the total recombination). The dark saturation current density $J_0$ is then given by: 
\begin{equation}
J_0=\frac{q}{\eta_{ext}} \int_0^\infty a(E) b_c(E) dE = \frac{q}{\eta_{ext}} \int_{E_g}^\infty b_c(E) dE.
\label{eq4}
\end{equation}

The current-voltage relationship for a photovoltaic cell is similar to Eqn. 7 in Ref. \cite{miller_strong_2012}, and is given as:
\begin{equation}
J(V)=J_{sc}-J_0 \exp \left( \frac{V}{k_B T_c} \right) = q \int_{E_g}^\infty b_s(E) dE - \frac{q}{\eta_{ext}} \exp \left( \frac{V}{k_B T_c} \right) \int_{E_g}^\infty b_c(E) dE,
\label{eq5}
\end{equation}
where $V$ is the voltage of the photovoltaic cell. In this analysis, we assume perfect carrier collection, i.e. at short circuit, every absorbed photon creates an electron-hole pair that is collected by the electrical cell contacts.

If we operate the photovoltaic cells at the maximum power point (the voltage $V_{MPP}$ at which the output power $P = J \times V$ is maximized), the output power is denoted as $J_{MPP} \times V_{MPP}$, where $J_{MPP}=J(V_{MPP})$. The thermophotovoltaic efficiency will be the ratio of output electrical power to input thermal power. The input thermal power equals the above-bandgap thermal radiation absorbed by the solar cells summed with the below-bandgap thermal radiation lost due to imperfect cell reflectivity, or equivalently, the total blackbody power incident on the cells minus the power that is reflected back to the hot radiation source.

The thermophotovoltaic efficiency $\eta_{ \hspace{.03cm} TPV}$ is thus given as:
\begin{equation}
\eta_{ \hspace{.03cm} TPV}=\frac{J_{MPP}V_{MPP}}{\int_0^\infty b_{s, \text{power}}(E) dE - R \int_0^{E_g} b_{s, \text{power}}(E) dE},
\label{eq6}
\end{equation}
where $R$ denotes the reflectivity of the photovoltaic cells to the sub-bandgap photons, modeled as a constant over all the sub-bandgap photon energies in this analysis, and $b_{s, \text{power}}$ is the blackbody spectrum in terms of power/area/energy. Any source of parasitic absorption of sub-bandgap photons, such as free carrier absorption, can be accounted for by penalizing $R$.

It should be noted that there is radiative emission out of the front of the cell, with the spectrum $R_{rad}(E)$ in units of power/area/energy: 
\begin{equation}
R_{rad}(E)=a(E) \frac{2 \pi E^3}{c^2 h^3 \left( \exp \left( \frac{E-qV}{k_B T_c} \right) - 1 \right)}.
\label{eq7}
\end{equation}
The integral of $R_{rad}(E)$ over photon energy $E$ is an additional amount of energy that is recycled back to the hot source. This term is ignored in this analysis, as its contribution is negligible.

\section{Results \& Discussion}
\begin{figure}[htbp]
  \centering
  \subfloat{\includegraphics[scale=0.45]{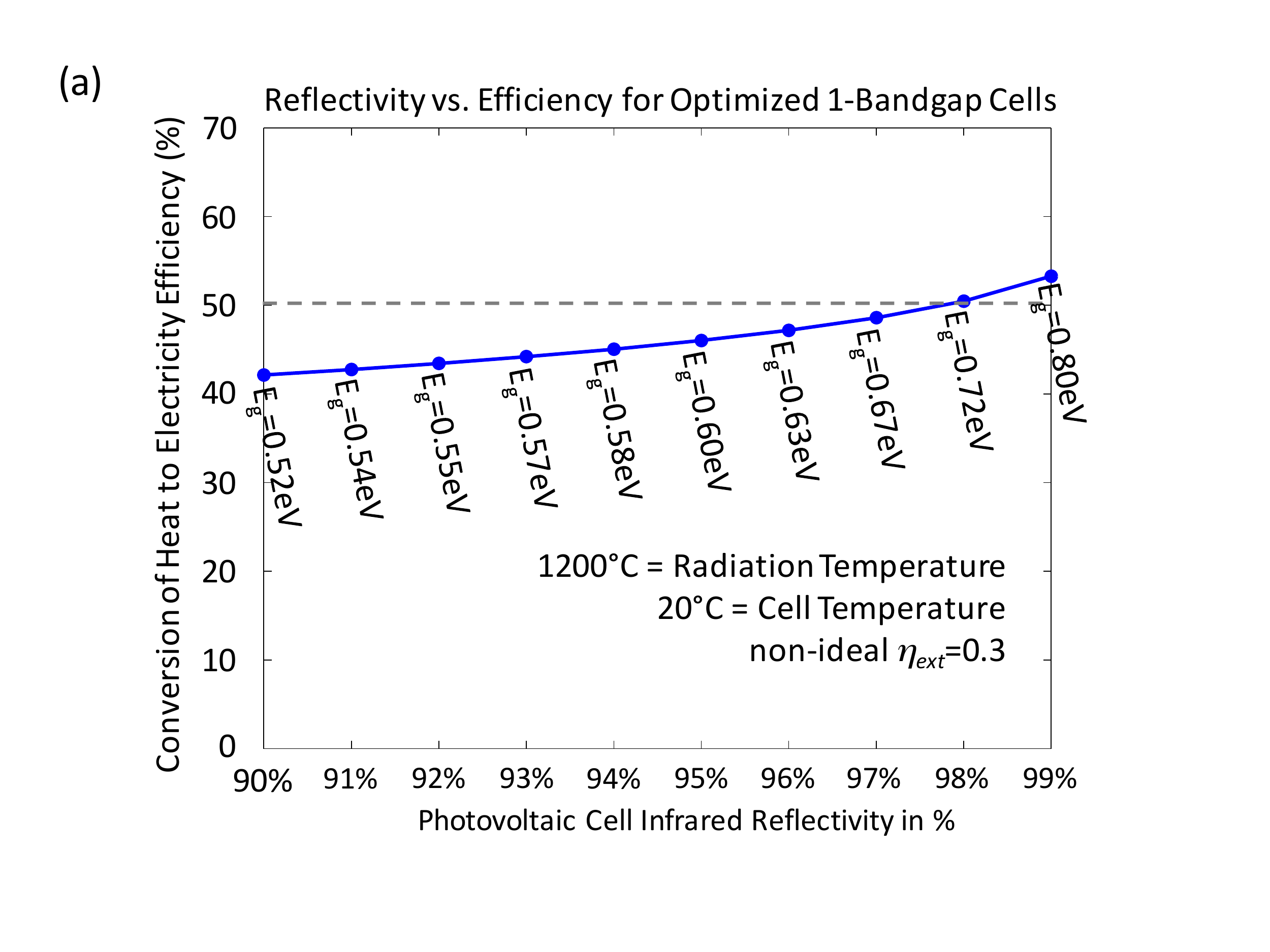}\label{fig4a}}
  \hfill
  \subfloat{\includegraphics[scale=0.45]{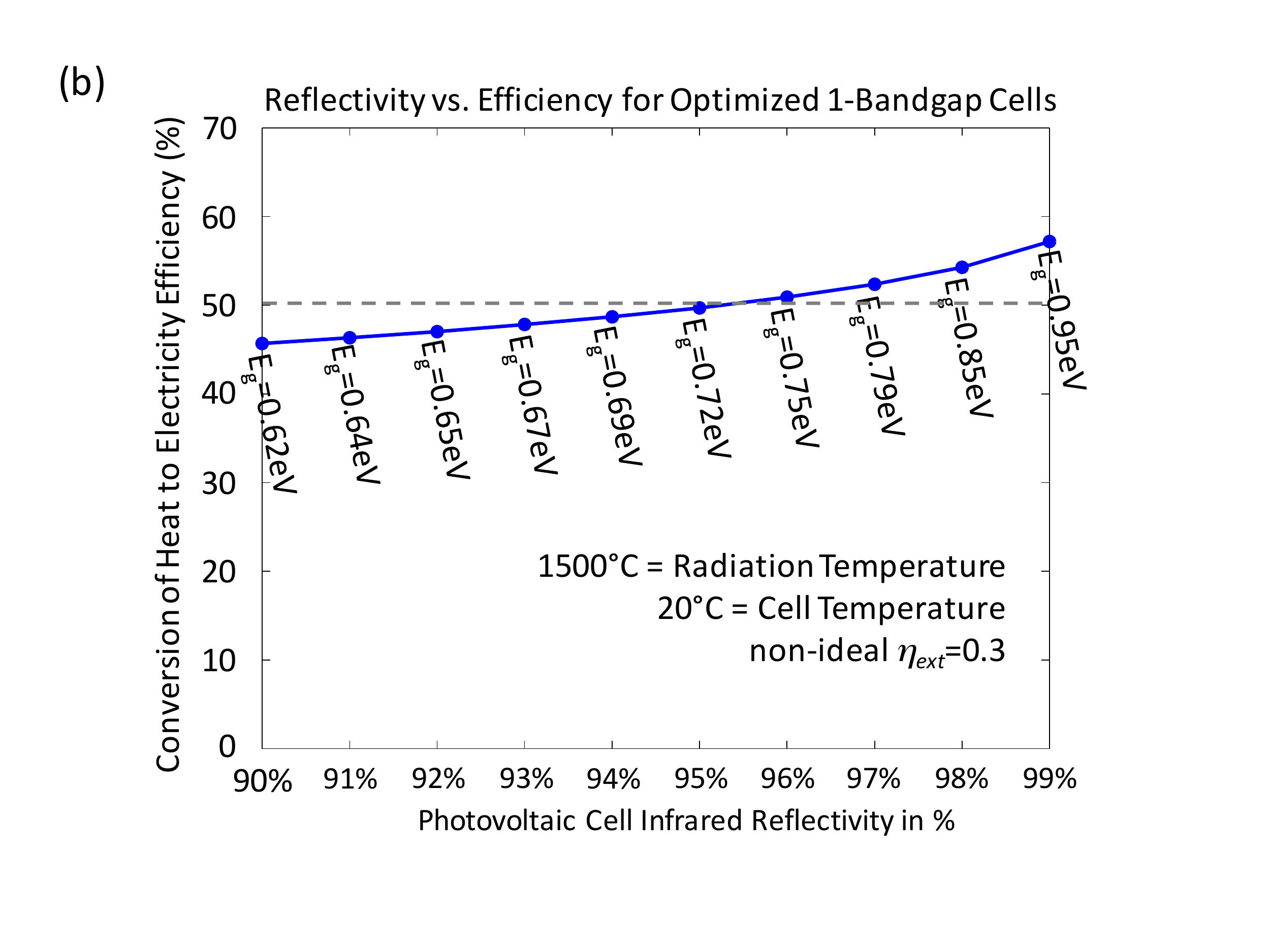}\label{fig4b}}
  \caption{The efficiency as a function of photovoltaic cell infrared reflectivity for single junction cells. The cell temperature is $20\degree$C, $\eta_{ext}=0.3$, and the source radiation temperature is (a) $1200\degree$C or (b) $1500\degree$C.}
  \label{fig4}
\end{figure}

We plot thermophotovoltaic efficiency $\eta_{ \hspace{.03cm} TPV}$ against the reflectivity $R$ of sub-bandgap photons  in Fig.~\ref{fig4}. For each value of $R$, we plot the cell bandgap $E_g$ that maximizes $\eta_{ \hspace{.03cm} TPV}$. We assume $T_c=20 \degree $ C in Fig.~\ref{fig4}; in Fig.~\ref{fig4a}, we assume a hot radiation source temperature $T_s=1200\degree$C, and in Fig.~\ref{fig4b}, we have $T_s=1500\degree$C. We do not assume a perfect photovoltaic cell, instead we set $\eta_{ext}=30\%$, the external luminescence yield in the record breaking Alta Devices solar cells. 

By plotting the optimal efficiency as a function of back mirror reflectivity, we show that we can compensate for a poor back mirror by reducing the bandgap of the photovoltaic cell. A poor back mirror means that more of the sub-bandgap photons will be lost. This loss can be ameliorated by moving to a smaller bandgap. In Fig.~\ref{fig4}, we mark 50\% efficiency by a gray dashed line. For $T_s=1200\degree$C in Fig.~\ref{fig4a}, we can achieve 50\% thermophotovoltaic efficiency with $R \geq 98\%$, and for $T_s=1500\degree$C, 50\% efficiency can be achieved with $R \geq 95\%$. At $1200\degree$C, a cell of bandgap 0.8 eV with $R=99\%$ can achieve 51\% thermophotovoltaic efficiency. At $1500\degree$C, a cell of bandgap 0.95 eV with $R=99\%$ can achieve 59\% thermophotovoltaic efficiency.

\begin{figure}[htbp]
\includegraphics[scale=0.5]{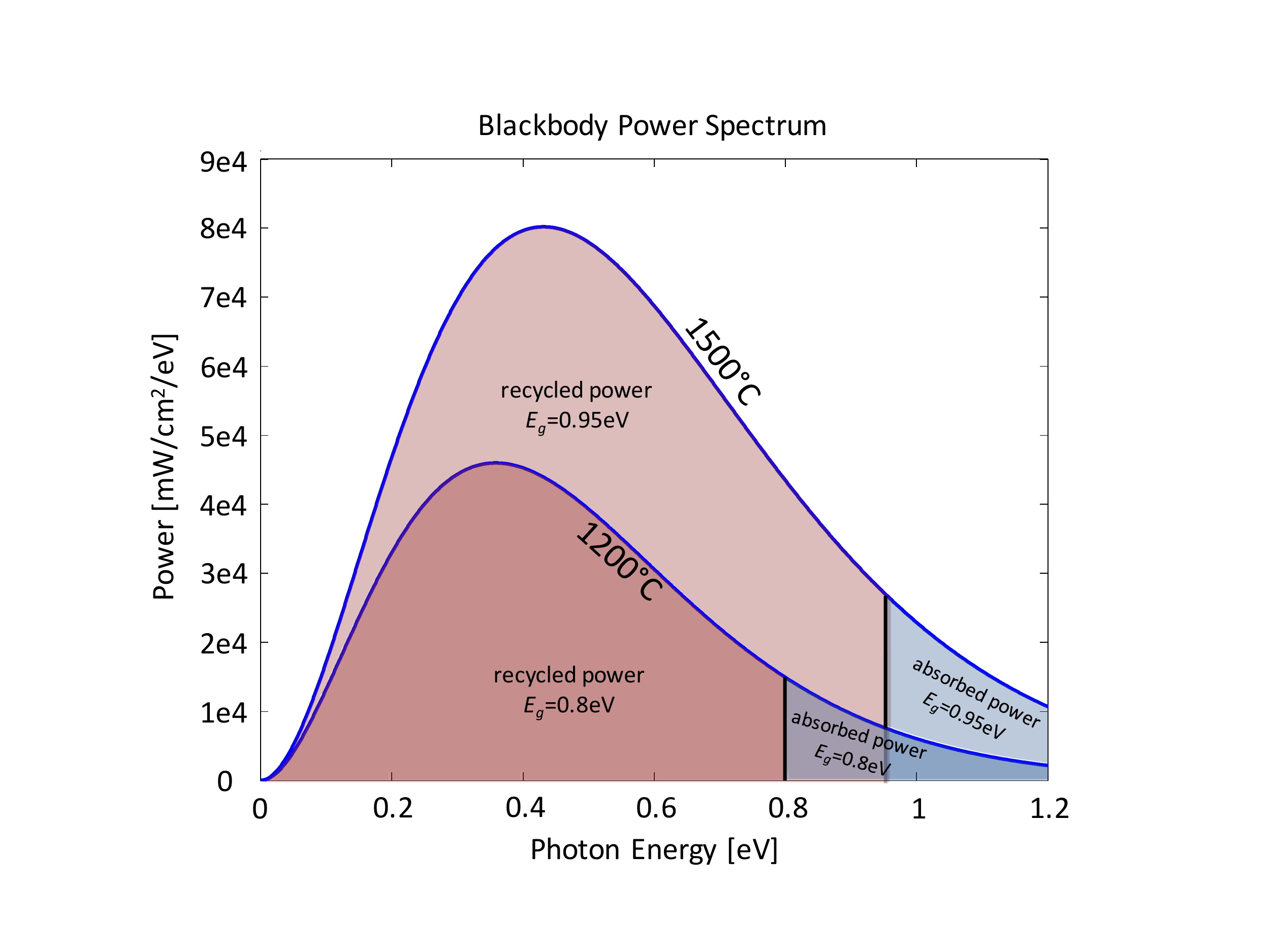}
\centering
\caption{The blackbody power spectrum, for blackbody sources at $1500\degree$C and $1200\degree$C. For a 0.8 eV bandgap cell and a $1200\degree$C source, the power absorbed by the solar cell is indicated by the blue region, and the power recycled back to the hot source is indicated by the red region. Similary for a 0.95 eV bandgap cell and a $1500\degree$C source, the regions of recycled and absorbed power are highlighted in red and blue, respectively.}
\label{fig5}
\end{figure}

Fig.~\ref{fig5} plots $b_{s, \text{power}}(E)$, the blackbody spectrum in terms of power/area/energy. For the bandgap of 0.8 eV at $1200\degree$C, we can integrate over the photon energies above 0.8 eV (the region highlighted in blue), to find the power/area that is absorbed by the 0.8 eV cell, and we can integrate over the photon energies below 0.8 eV (the region highlighted in red) to find the sub-bandgap power per unit area that can potentially be reflected back to the source. At $1200\degree$C for a 0.8 eV bandgap, there is 27 W/cm$^2$ of black body radiation, of which 3.2 W/cm$^2$ is absorbed, and 1.56 W/cm$^2$ is converted to electricity. Thus, up to $\approx 24$ W/cm$^2$ is re-thermalized on each reflection. The large percentage of sub-bandgap power highlights the need for high reflectivity back to the hot source.

We see that a reflective mirror on the backside of the photovoltaic cell has great importance to the thermophotovoltaic efficiency. The question then arises: can we obtain further benefit from using a reflective back mirror and a selective source?  Let us say that our selective source has a sub-bandgap reflectivity $R_s$, and the sub-bandgap reflectivity of the cell back mirror is $R_c$. A sub-bandgap photon then sees an effective reflectivity $R$, given by the infinite summation: 
\begin{eqnarray}
 R &=& R_s + (1-R_s) R_c (1-R_s)+(1-R_s) R_c R_s R_c (1-R_s) \nonumber \\
      &&              + (1-R_s) R_c R_s R_c R_s R_c (1-R_s) \ldots      \nonumber \\
   &=& R_s + (1-R_s)^2 R_c \sum_{n=0}^{\infty} R_c^n R_s^n \nonumber \\
   &=& \frac{R_s+R_c - 2 R_s R_c}{1 - R_c R_s}.
\label{eq8}   
\end{eqnarray}
If we have $R_s \gg R_c$, the equation reduces down to $\approx R_s$. Likewise, if $R_c \gg R_s$, the equation reduces down to $\approx R_c$. In these cases, there is not much additional benefit from having both a selective source and a reflective back mirror; the one with higher reflectivity dominates. Even if $R_c \approx R_s$, there is diminishing additional benefit as the reflectivity goes higher.

We have assumed in our analysis a thermophotovoltaic cell with an external luminescence efficiency of $\eta_{ext}=30\%$, a realistic value for high-quality GaAs. However, though ideally suited for solar energy conversion, GaAs has too large a bandgap to be suitable for thermophotovoltaics. Is there a material within the bandgap ranges shown in Fig. \ref{fig4} that can reach a similar value of external luminescence efficiency?

For radiator temperatures in the 1200$\degree$C to 1500$\degree$C range, $\text{In}_{0.53}\text{Ga}_{0.47}\text{As}$ ($E_g = 0.75$ eV) is a candidate material, provided that the back surface has a reflectivity of at least 98\%. This material can be grown lattice-matched to an InP substrate, a thin film of which can be separated from the substrate by an epitaxial lift-off process. The external luminescence efficiency of the device is given by:
\begin{equation}
\eta_{ext}=\frac{\exp \left( \frac{qV}{k_B T_c} \right) \int_0^\infty a(E) b_c(E) dE}{\exp \left( \frac{qV}{k_B T_c}\right) \int_0^\infty a(E) b_c(E) dE + \Phi_{back} + \Phi_{nr}},
\label{eq9}
\end{equation}
where the numerator is the emitted flux out of the front surface, and the denominator also includes the rate of loss per unit device area through back absorption $\Phi_{back}$ and through non-radiative recombination $\Phi_{nr}$.

To accurately calculate $\eta_{ext}$, we cannot assume an ideal step function absorption profile. In a planar cell, the photon has two passes through the cell to be absorbed, with a chance of loss at the back surface if the back reflectivity $R < 100\%$. The absorptivity is:
\begin{equation}
a(E) = 1 - e^{-\alpha(E)L} + Re^{-\alpha(E)L}\left( 1 - e^{-\alpha(E)L} \right)
\end{equation}
where $L$ is the cell thickness, $\alpha (E)$ is the material's absorption coefficient, and we have assumed a transparent front surface. We have neglected the very weak angle dependence of $a(E)$ since all of the incident rays are refracted into a narrow cone near the normal when they enter the high-index semiconductor. We model the absorption coefficient of $\text{In}_{0.53}\text{Ga}_{0.47}\text{As}$ with a piece-wise fit:
\begin{equation}
    \alpha (E) = 
\begin{dcases}
     \alpha_0 \exp \left( \frac{E - E_g}{E_0} \right),& \text{if } E<E_g\\
    \alpha_0 \left( 1+ \frac{E-E_g}{E'}\right),              &  \text{if } E \geq E_g
\end{dcases}
\label{eq13}
\end{equation}
Similarly to \cite{jurczak_efficiency_2015}, we obtain the band-edge absorption $\alpha_0$ and Urbach tail energy $E_0$ for $\text{In}_{0.53}\text{Ga}_{0.47}\text{As}$ by a linear interpolation of the corresponding values for GaAs ($\alpha_0 = 8000$ cm$^{-1}$, $E_0 = 6.7$ meV) \cite{sturge_optical_1962} and InAs ($\alpha_0 = 2500$ cm$^{-1}$, $E_0 = 6.9$ meV) \cite{dixon_optical_1961}, based on composition. An above-bandgap parameter $E' = 140$ meV was used, which matches the absorption properties of GaAs \cite{sturge_optical_1962}. We assume the refractive index of $\text{In}_{0.53}\text{Ga}_{0.47}\text{As}$ to be approximately the same as for GaAs, with $n_r  = 3.5$.

To calculate the back absorption rate $\Phi_{back}$, we treat the back surface as a partially absorbing substrate with the same refractive index as $\text{In}_{0.53}\text{Ga}_{0.47}\text{As}$. The absorption of luminescent photons by the back is exactly balanced by the thermal radiation, enhanced by $\exp(qV/k_BT_c)$, that enters from the back and is subsequently absorbed by the semiconductor. The result is:
\begin{equation}
\Phi_{back}= \exp \left( \frac{qV}{k_B T_c} \right) \cdot 2\pi \int_0^\infty \int_0^\frac{\pi}{2} a_{back}(E,\theta) \, \frac{2n_r^2 E^2}{c^2 h^3 } \exp\left(-\frac{E}{k_BT_c} \right) \, \cos \theta \sin \theta \, d\theta \, dE
\label{eq10}
\end{equation}
where $a_{back}$ is the absorptivity of the cell to photons incident from the back, which depends strongly on both the angle $\theta$ and energy $E$. These photons first see a transmissivity through the back surface of $(1 - R)$ prior to entering the semiconductor. Afterwards, if the photon lies inside the front escape cone, it only has one pass through the cell to be absorbed. If the photon lies outside the escape cone, it is totally reflected from the front surface and partially reflected from the back surface, and has many opportunities to be absorbed after multiple reflections. The angle-dependent back absorptivity is therefore given by:
\begin{equation}
a_{back}(E,\theta) = (1-R) \times
\begin{dcases}
     1 - \exp \left(- \frac{\alpha(E)L}{\cos \theta} \right), & \text{if } \theta<\theta_c\\
     \frac{ 1 - \exp \left(- \frac{2\alpha(E)L}{\cos \theta} \right) } {1 - R\exp \left(- \frac{2\alpha(E)L}{\cos \theta} \right) }, &  \text{if } \theta \geq \theta_c
\end{dcases}
\label{eq11}
\end{equation}
where $\theta_c = \sin^{-1}\left(1/n_r\right)$ is the critical angle of the front interface, and we have assumed for simplicity that the back reflectivity $R$ is independent of angle and energy. For this calculation, we assume a fixed reflectivity of $R = 98\%$ both for the luminescent photons and for the unabsorbed photons below the bandgap. In practice, these reflectivities may not be identical.

Non-radiative recombination occurs through Shockley-Read-Hall (SRH) and Auger recombination. The total recombination rate for these processes is given by:
\begin{equation}
\Phi_{nr} = L \times \left[ \frac{np - n_i^2}{\tau_{SRH} \cdot \left(n+p+2n_i \right)} + \left(C_n n + C_p p \right) \cdot (n p - n_i^2) \right]
\label{eq12}
\end{equation}
where $n$ and $p$ are the free electron and hole concentrations, respectively, $n_i$ is the intrinsic carrier density, $\tau_{SRH}$ is the SRH recombination lifetime (assumed equal for both carriers), and $C_n$ and $C_p$ are the Auger recombination coefficients for the two-electron and two-hole processes, respectively. We assume an $\text{In}_{0.53}\text{Ga}_{0.47}\text{As}$ cell operated at a temperature of $T_c = 20\degree$C with an $n$-type doping density of $N_D = 1 \times 10^{17}\text{ cm}^{-3}$. We use the electronic parameters for $\text{In}_{0.53}\text{Ga}_{0.47}\text{As}$ given in \cite{nsm} to evaluate the carrier densities as a function of the voltage on the cell, which is under illumination from the thermal emitter. In a high-quality film, the SRH lifetime of the minority carriers can be up to $\tau_{SRH} = 47.36$ $\mu$s and the Auger coefficients are $C_n = C_p =8.1 \times 10^{-29}$ cm$^{-6}$s$^{-1}$ \cite{ahrenkiel_recombination_1998}.

Decreasing the cell thickness $L$ reduces the rate of non-radiative recombination, as shown by Eqn. 14. However, this comes at a cost to the front absorptivity $a(E)$, leading to diminished external emission as well as a lower short-circuit current. Balancing these design trade-offs, we obtain an optimal thickness of $L  = 1.5$ $\mu$m for the thermophotovoltaic cell with a 98\% reflective back surface. This device achieves an external luminescence efficiency of $\eta_{ext} = 31.9\%$ at open-circuit, which validates our prior assumption of $\eta_{ext} = 30\%$.

Meanwhile, this thickness still allows for near-complete absorption of above-bandgap photons from an emitter at $T_s = 1200\degree$C, resulting in a short-circuit current $J_{sc}$ that is 97\% of the value for an ideal step-function absorber. At this temperature, the system attains a thermophotovoltaic conversion efficiency of $\eta_{TPV} = 49.9\%$. This is close to the limit shown in Fig. \ref{fig4}, which suggests that $\text{In}_{0.53}\text{Ga}_{0.47}\text{As}$ is a prime material for high-performance thermophotovoltaics.

\section{Conclusion}
Thermophotovoltaics has the potential to be a highly efficient method of heat to electricity conversion that is also portable and compact, containing no moving parts. The idea of thermophotovoltaics was established in 1956 \cite{nelson_brief_2003}, though at that point in time, photovoltaic cells, especially low-bandgap cells, were too inefficient for the idea to take off. More recently, efforts have focused around designing a photonic crystal to be a selective hot emitter \cite{gee_selective_2002}. The photonic crystal is engineered to suppress emission of the photons with energy below the photovoltaic bandgap, while allowing emission of photons with energy above the bandgap. Designing a photonic crystal with emissivity matching the absorptivity of the photovoltaic cell is a difficult challenge; in a recent effort, the emissivity of below-bandgap photons in a tantalum 2D photonic crystal emitter was $\approx 30\%$ in simulation (analogous to only 70\% reflectivity of sub-bandgap photons) \cite{yeng_performance_2013}. A photonic crystal also has reliability problems at high temperatures due to its nano- and micro-structuring. Finding a bulk material with the desired emissivity is also a challenge; it has also been proposed to use a bulk refractory metal such as titanium nitride (TiN) for the hot source \cite{guler_refractory_2014}, but TiN still has an emissivity of $\approx 30\%$ for low energy infrared photons \cite{naik_titanium_2012}. 

Exploiting the photovoltaic cell band-edge itself as the spectral filter eliminates the difficult problem of aligning the source emissivity to the cell absorptivity; the alignment becomes automatic. The serendipitous development of a photovoltaic cell back mirror that reflects sub-bandgap radiation has provided a breakthrough for thermophotovoltaics. Without special attention to the sub-bandgap photon reflectivity, a standard production Alta Devices solar cell reached $R>92\%$, and the recording breaking GaAs cell had $~98\%$ sub-bandgap photon reflectivity. $R>99\%$ is achievable. With a $1200\degree$C source, if all photons above $\approx 0.8$ eV can be used, and 99\% of unabsorbed photons below $\approx 0.8$ eV energy can be recycled to the heat source, the conversion from heat to electricity can be $>50\%$ efficient. We analyze the case of a photovoltaic cell with one material bandgap in this work, but even higher efficiencies may be achieved if photovoltaic cells with multiple bandgaps are used.

{\footnotesize
\bibliography{ThermoPhotoVoltaics}{}

\begin{thebibliography}{10}

\bibitem{fraas_tpv_2007}
L.~Fraas and L.~Minkin, ``{TPV} {History} from 1990 to {Present} \& {Future}
  {Trends},'' {\em AIP Conference Proceedings}, vol.~890, pp.~17--23, 2007.

\bibitem{harder_theoretical_2003}
N.-P. Harder and P.~Wurfel, ``Theoretical limits of thermophotovoltaic solar
  energy conversion,'' {\em Semiconductor Science and Technology}, vol.~18,
  pp.~S151--S157, May 2003.

\bibitem{teofilo_thermophotovoltaic_2008}
V.~L. Teofilo, P.~Choong, J.~Chang, Y.-L. Tseng, and S.~Ermer,
  ``Thermophotovoltaic {Energy} {Conversion} for {Space},'' {\em The Journal of
  Physical Chemistry C}, vol.~112, pp.~7841--7845, May 2008.

\bibitem{werth_thermo-photovoltaic_1967}
J.~Werth, ``Thermo-photovoltaic converter with radiant energy reflective
  means,'' July 1967.
\newblock US Patent 3,331,707.

\bibitem{miller_strong_2012}
O.~D. Miller, E.~Yablonovitch, and S.~R. Kurtz, ``Strong {Internal} and
  {External} {Luminescence} as {Solar} {Cells} {Approach} the
  {Shockley}-{Queisser} {Limit},'' {\em IEEE Journal of Photovoltaics}, vol.~2,
  pp.~303--311, July 2012.

\bibitem{green_solar_2011-1}
M.~A. Green, K.~Emery, Y.~Hishikawa, and W.~Warta, ``Solar cell efficiency
  tables (version 37),'' {\em Progress in Photovoltaics: Research and
  Applications}, vol.~19, pp.~84--92, Jan. 2011.

\bibitem{green_solar_2011}
M.~A. Green, K.~Emery, Y.~Hishikawa, W.~Warta, and E.~D. Dunlop, ``Solar cell
  efficiency tables ({version} 38),'' {\em Progress in Photovoltaics: Research
  and Applications}, vol.~19, pp.~565--572, Aug. 2011.

\bibitem{green_solar_2012}
M.~A. Green, K.~Emery, Y.~Hishikawa, W.~Warta, and E.~D. Dunlop, ``Solar cell
  efficiency tables (version 39),'' {\em Progress in Photovoltaics: Research
  and Applications}, vol.~20, pp.~12--20, Jan. 2012.

\bibitem{green_solar_2012-1}
M.~A. Green, K.~Emery, Y.~Hishikawa, W.~Warta, and E.~D. Dunlop, ``Solar cell
  efficiency tables (version 40),'' {\em Progress in Photovoltaics: Research
  and Applications}, vol.~20, pp.~606--614, Aug. 2012.

\bibitem{swanson_silicon_1978}
R.~Swanson, ``Silicon photovoltaic cells in thermophotovoltaic energy
  conversion,'' {\em International Electron Devices Meeting}, pp.~70--73, 1978.

\bibitem{siergiej_20_2003}
R.~R. Siergiej, B.~Wernsman, S.~A. Derry, R.~G. Mahorter, R.~J. Wehrer, S.~D.
  Link, M.~N. Palmisiano, R.~L. Messham, S.~Murray, C.~S. Murray, F.~Newman,
  J.~Hills, and D.~Taylor, ``20\% {Efficient} {InGaAs}/{InPAs}
  {Thermophotovoltaic} {Cells},'' {\em AIP Conference Proceedings}, vol.~653,
  pp.~414--423, 2003.

\bibitem{wernsman_greater_2004}
B.~Wernsman, R.~Siergiej, S.~Link, R.~Mahorter, M.~Palmisiano, R.~Wehrer,
  R.~Schultz, G.~Schmuck, R.~Messham, S.~Murray, C.~Murray, F.~Newman,
  D.~Taylor, D.~DePoy, and T.~Rahmlow, ``Greater {Than} 20\% {Radiant} {Heat}
  {Conversion} {Efficiency} of a {Thermophotovoltaic} {Radiator}/{Module}
  {System} {Using} {Reflective} {Spectral} {Control},'' {\em IEEE Transactions
  on Electron Devices}, vol.~51, pp.~512--515, Mar. 2004.

\bibitem{yablonovitch_extreme_1987}
E.~Yablonovitch, T.~Gmitter, J.~P. Harbison, and R.~Bhat, ``Extreme selectivity
  in the lift-off of epitaxial {GaAs} films,'' {\em Applied Physics Letters},
  vol.~51, no.~26, p.~2222, 1987.

\bibitem{kayes_27.6_2011}
B.~M. Kayes, H.~Nie, R.~Twist, S.~G. Spruytte, F.~Reinhardt, I.~C. Kizilyalli,
  and G.~S. Higashi, ``27.6\% {Conversion} efficiency, a new record for
  single-junction solar cells under 1 sun illumination,'' {\em 37th IEEE
  Photovoltaic Specialists Conference}, pp.~4--8, June 2011.

\bibitem{shockley_detailed_1961}
W.~Shockley and H.~J. Queisser, ``Detailed {Balance} {Limit} of {Efficiency} of
  p-n {Junction} {Solar} {Cells},'' {\em Journal of Applied Physics}, vol.~32,
  pp.~510--519, Mar. 1961.

\bibitem{nsm}
``{NSM} {Archive} - {Gallium} {Indium} {Arsenide} ({GaInAs}) - {Band}
  structure.''
\newblock [Online: accessed 1/7/2016].

\bibitem{ahrenkiel_recombination_1998}
R.~K. Ahrenkiel, R.~Ellingson, S.~Johnston, and M.~Wanlass, ``Recombination
  lifetime of $\text{In}_{0.53}\text{Ga}_{0.47}\text{As}$ as a function of
  doping density,'' {\em Applied Physics Letters}, vol.~72, no.~26, p.~3470,
  1998.

\bibitem{jurczak_efficiency_2015}
P.~Jurczak, A.~Onno, K.~Sablon, and H.~Liu, ``Efficiency of {GaInAs}
  thermophotovoltaic cells: the effects of incident radiation, light trapping
  and recombinations,'' {\em Optics Express}, vol.~23, p.~A1208, Sept. 2015.

\bibitem{sturge_optical_1962}
M.~D. Sturge, ``Optical {Absorption} of {Gallium} {Arsenide} between 0.6 and
  2.75 {eV},'' {\em Physical Review}, vol.~127, pp.~768--773, Aug. 1962.

\bibitem{dixon_optical_1961}
J.~R. Dixon and J.~M. Ellis, ``Optical {Properties} of n-{Type} {Indium}
  {Arsenide} in the {Fundamental} {Absorption} {Edge} {Region},'' {\em Physical
  Review}, vol.~123, pp.~1560--1566, Sept. 1961.

\bibitem{nelson_brief_2003}
R.~E. Nelson, ``A brief history of thermophotovoltaic development,'' {\em
  Semiconductor Science and Technology}, vol.~18, pp.~S141--S143, May 2003.

\bibitem{gee_selective_2002}
J.~Gee, J.~Moreno, {Shawn-Yu Lin}, and J.~Fleming, ``Selective emitters using
  photonic crystals for thermophotovoltaic energy conversion,'' {\em 29th IEEE
  Photovoltaic Specialists Conference}, pp.~896--899, 2002.

\bibitem{yeng_performance_2013}
Y.~X. Yeng, W.~R. Chan, V.~Rinnerbauer, J.~D. Joannopoulos, M.~Soljacic, and
  I.~Celanovic, ``Performance analysis of experimentally viable photonic
  crystal enhanced thermophotovoltaic systems,'' {\em Optics Express}, vol.~21,
  p.~A1035, Nov. 2013.

\bibitem{guler_refractory_2014}
U.~Guler, A.~Boltasseva, and V.~M. Shalaev, ``Refractory {Plasmonics},'' {\em
  Science}, vol.~344, pp.~263--264, Apr. 2014.

\bibitem{naik_titanium_2012}
G.~V. Naik, J.~L. Schroeder, X.~Ni, A.~V. Kildishev, T.~D. Sands, and
  A.~Boltasseva, ``Titanium nitride as a plasmonic material for visible and
  near-infrared wavelengths,'' {\em Optical Materials Express}, vol.~2, p.~478,
  Apr. 2012.

\end{thebibliography}
\bibliographystyle{ieeetr}}
\end{document}